\documentclass{elsart}
\usepackage{natbib}

\usepackage{epsf}

\newcommand{\lesssim}{\lower.5ex\hbox{$\; \buildrel < \over\sim \;$}}
\newcommand{\gtrsim}{\lower.5ex\hbox{$\; \buildrel > \over\sim \;$}}

\begin{document}
\runauthor{Dermer and Dingus}
\begin{frontmatter}
\title{Blazar Flaring Rates Measured with {\it GLAST}}
\author[NRL]{C.\ D.\ Dermer,}
\author[LANL]{B.\ L.\ Dingus}

\address[NRL]{Code 7653, Naval Research Laboratory, Washington, DC 20375-5352 U.S.A.}
\address[LANL]{Los Alamos National Laboratory, MS H803 P-23, Los Alamos, NM 87545 U.S.A.}

\begin{abstract}
We derive the minimum observing time scales to detect a blazar at a
given flux level with the LAT on {\it GLAST} in the scanning and
pointing modes.  Based upon Phase 1 observations with EGRET, we
predict the {\it GLAST} detection rate of blazar flares at different
flux levels. With some uncertainty given the poor statistics of bright
blazars, we predict that a blazar flare with integral flux $\gtrsim
200 \times 10^{-8}$ ph($> 100$ MeV) cm$^{-2}$ s$^{-1}$, which are the
best candidates for Target of Opportunity pointings and extensive
temporal and spectral studies, should occur every few days.
\end{abstract}
\end{frontmatter}

\section{Introduction}

The statistical significance for detection of a known source
at the $n\sigma$ level is $n \cong S/ \sqrt{B}$,
where $S$ is the number of source counts, $B$ is the number of
background counts, and the diffuse background is assumed to be
precisely known \cite{lm83}. The observer is also assumed to follow a
predetermined data analysis protocol and not to make multiple
rebinnings of the data.

In this paper, we derive the minimum observing time for the Large Area Telescope
(LAT) on {\it GLAST} (Gamma ray Large Area
Space Telescope) to detect
high-latitude $\gamma$-ray sources at given flux levels, and estimate
the rate at which blazar flares at certain flux levels are expected
to be observed with {\it GLAST}.

In Sections 2 and 3, we present calculations of signal and background
counts, respectively. The five count and the five sigma requirements
for detection are derived in Section 4. EGRET blazar studies,
described in Section 5, are used to infer the probable rate of flaring
emissions that GLAST will measure, as discussed in Section 6.

\section{Source Counts $S$}

The point-spread function (psf) of the LAT on {\it GLAST} is
characterized by angle
$\theta_{psf,G}=$ $(3.5^\circ/57.3^\circ)$ $(E/E_{100})^{-2/3}\equiv
0.061u^{-2/3}$, where $E_{100} = 100$ MeV and $E=E$(MeV)
\cite{glast}. We assume an energy-dependent analysis of the {\it
GLAST} data, where the number of counts is integrated in the direction
of a catalogued source over the solid angle $\Delta\Omega_G = \pi
\theta_{psf,G}^2 = 0.012 u^{-4/3}$. For this analysis procedure,
the fraction $f_\gamma$ of detected source photons is therefore
independent of energy. For an azimuthally-symmetric, Gaussian
distribution $\propto \exp [-(\theta/\theta_{psf,G})^2]$ of count
directions, $f_\gamma = 1-\exp(-1) = 0.63$, though $f_\gamma\cong
0.68$ would apply if $\theta_{psf}$ is defined as the 68\% containment
angle.

The number of source photon counts with energy $> E$ within $\Delta
\Omega_G$ of a given point source is, in this approximation,given by
\begin{equation}
S \cong f_\gamma \int _0^{\Delta t} dt \int_{E}^{\infty} 
dE^\prime \; A[E^\prime,\theta(t),\phi(t)]\; \phi_s(E^\prime,t)
\label{S}
\end{equation}
where $\phi_s(E,t)$ is the source photon flux (ph cm$^{-2}$ s$^{-1}$
E$^{-1}$). We approximate the energy- and angle-dependent effective
area of the LAT by the function
\begin{equation}
A(E) = A_0 u(\theta)(E/E_{100})^{a(\theta)}\cong X A_0 u^{a_0} \;,
\label{A(E)}
\end{equation}
where $u(\theta = 0^\circ) = 1$, $a(\theta = 0^\circ) = a_0$, and
$\theta$ is the angle between the direction of a source and the
axis of the LAT, and $\phi$ gives the azimuthal angle of the
source with respect to the LAT orientation. Here we
assume azimuthal symmetry of the detector effective area.  An
approximation for the effective area that satisfies the {\it GLAST} Science
Requirements Document is $A_0 \approx 6200$ cm$^2$ and $a_0
\approx 0.16$ for 100 MeV $\lesssim E \lesssim 10$ GeV 
(compare Foldout B in the {\it GLAST}
proposal; this approximation breaks down above 10 GeV). 
For 20 MeV $\lesssim E
\lesssim 100$ MeV, $a_0 \approx 0.37$. The net
observing time is denoted by $\Delta t=10^4 t_4$ s, and the exposure
factor X accounts for the fraction of time that a given source is
being detected by the LAT. The effective area of the LAT drops by a
factor-of-two at $\theta = 55^\circ$, so that we take $X =0.2 X_{0.2}$
with $X_{0.2} \cong 1$ in the scanning mode, noting that ${1\over
2}(1-\cos 55^\circ )=$21\%. (Signal detection on a $10^4$ second time
scale is of special interest in blazar studies, because the
light-travel time across the Schwarzschild radius of a $10^9 M_\odot$
is $\simeq 10^4$ s.)

Normalizing the source flux to $10^{-8}\phi_{-8}~{\rm ph}(>100$ MeV)
cm$^{-2}$ s$^{-1}$, we have $\phi_s(E,t) = ku^{-\alpha_{ph}}$, where
$k = 10^{-8}\alpha \phi_{-8}/E_{100}$. For $E> 100$ MeV,
\begin{equation}
S \cong {0.15\over g}\;f_{\gamma} \;(X_{0.2}t_4) 
\alpha\phi_{-8} u^{\alpha_\nu-0.84}\;,
\label{S1}
\end{equation}
where $g = g(\alpha_\nu) = 1-(\alpha_\nu/0.84)$. Here we use the
following conventions to define the photon number index $\alpha_{ph}$,
energy index $\alpha$, and $\nu F_\nu$ index $\alpha_\nu$:
$\alpha=\alpha_{ph} -1$ and $\alpha_\nu = 1-\alpha = 2-\alpha_{ph}$.
For an intrinsic source spectrum with photon index $\alpha_{ph}\cong
2$, most of the photons are detected near the lower energy threshold.

\section{Background Counts}

The number of background photons with energies $>E$ observed during
time $\Delta t$ that lie within solid element $\Delta \Omega$ that is
centered in the direction $\vec\Omega$ of a specified source is
\begin{equation}
B \cong \int_0^{\Delta t}dt\int_{E}^\infty dE^\prime \;
\Delta \Omega(E^\prime) \; A[E^\prime,\theta(t),
\phi(t)]\;\Phi_B(E^\prime,\vec\Omega)\; ,
\label{B}
\end{equation}
where $\Phi_B(E^\prime,\vec\Omega)$ is the diffuse $\gamma$-ray
background in the source direction.  A minimum level of background is
provided by the diffuse extragalactic $\gamma$-ray background
radiation, given by
\begin{equation}
\Phi_X(E) = k_x u^{-\alpha_B}\; , 0.4\lesssim u \lesssim 700
\label{Phi_B}
\end{equation}
\cite{sre98}, where $k_x = 1.73\pm 0.08 \times 10^{-7}$ ph
(cm$^2$-s-sr-MeV)$^{-1}$ and $\alpha_B = 2.10\pm 0.03$.  Other sources
of background, in particular, the galactic diffuse $\gamma$-ray
background, can also contribute; the $>100$ MeV photon flux is
$\approx$ twice as bright at $|b|=35^\circ$ and over 30 times brighter
at the Galactic center than at the poles \cite{hks97}. A better estimate
of blazar flaring rates must consider the galactic diffuse emission.

Considering only the extragalactic diffuse radiation and letting
$\Delta\Omega = \Delta\Omega_G$, the number of
background counts detected by {\it GLAST} is
\begin{equation}
B \cong 1.1 X_{0.2}t_4 u^{-2.27}\;.
\label{B1}
\end{equation}
This expression is valid for $1 \lesssim u \lesssim 100$.

\section{Five Count and 5$\sigma$ Criteria}

The detection of at least 5 counts implies from equation (4) that $S >
5S_5$ with $S_5 \geq 1$ (the 5 count criterion), so that
\begin{equation}
(X_{0.2}t_4) \phi_{-8} \gtrsim {34\over \alpha 
f_\gamma}\; gS_5 u^{0.84-\alpha_\nu}\;.\;
\label{S5}
\end{equation}
This expression fails for hard spectrum sources with $\alpha_\nu
\gtrsim 0.84$, where cutoffs or breaks in the higher energy range of
the spectrum and high-energy corrections to the {\it GLAST} effective
area determine the number of detected source counts. Otherwise, most
of the source counts are detected at the lower energy range near $E
\cong E_{100}$. At energies below 100 MeV, the effective area declines
more rapidly, so that most photons will be detected at even lower
energies than 100 MeV unless $\alpha_\nu \gtrsim 0.63$.

Detection of a source with significance $> 5 \sigma$ requires that
$S/\sqrt{B} > 5\sigma_5$. Using expressions (4) and (7) for
$S$ and $B$, respectively, gives
\begin{equation}
(X_{0.2}t_4) \phi_{-8} \gtrsim {1220\over \phi_{-8}} 
\; ({g \sigma_5\over \alpha f_\gamma})^2\; u^{-2(\alpha_\nu +0.3)}\;.\;
\label{sigma1}
\end{equation}
By comparing equations (8) and (9), one sees that detection of
soft-spectrum ($\alpha_\nu \lesssim 0$) sources with the LAT at the
level of $\phi_{-8} \cong 15$, corresponding to the minimum flux
sources that EGRET could detect in a 2 week viewing period, are
background-limited near 100 MeV (that is, the time-on-source
$X_{0.2}t_4$ is larger for the $5 \sigma$ criterion when $u \approx 1$
than for the 5 count criterion).

For spectra that are not very soft (i.e., $\alpha_\nu \gtrsim -0.3$),
the significance of detection increases with energy even as the source
counting rate declines. Therefore there is a unique energy $\bar E =
E_{100} \bar u$ at which both the 5 count and $5 \sigma$ criteria for
a source emitting at a given flux level $\phi_{-8}$ are satisfied, and
this energy defines the minimum observing time to satisfy these
criteria.  For sources with $\alpha_\nu \gtrsim -0.3$, this energy is
\begin{equation}
\bar u \cong ({36 g\sigma_5^2 \over 
\alpha f_\gamma \phi_{-8}S_5})^{1/(1.43+\alpha_\nu)}
\rightarrow 16\;{\sigma_5^{1.4}\over (\phi_{-8} S_5)^{0.7}}\;,
\label{baru}
\end{equation}
where the last expression applies in the special case $\alpha_\nu = 0,
f_\gamma = 0.68$. When $\phi_{-8} = 15$, the energy to meet the two
criteria is $\bar u \cong 4$, or the greatest sensitivity is at
$\approx 200$-300 MeV.  For sources softer (more negative) than
$\alpha_\nu \cong -0.3$, energies near 100 MeV give the best sensitivity.

The minimum observing time is given by
\begin{equation}
( X_{0.2} t_4 )_{min} = {34\over \alpha f_\gamma}\; 
{gS_5\over \phi_{-8}}\; \bar u^{~0.84-\alpha_\nu}\rightarrow 
7 \;S_5^{0.41} \sigma_5^{1.18} ({\phi_{-8}\over 15})^{-1.59}\;,
\label{<Xt>}
\end{equation}
where the right-hand-side is evaluated in the special case described
above. Thus, {\it GLAST} will detect high-latitude sources at the
level EGRET could reach in two weeks in only $\sim 7\times 10^4$ s
$\cong 0.8$ days, and it accomplishes this task in the scanning
mode. Considering a background twice as large, which is more realistic
for the majority of blazars, gives a detection of $\phi_{-8} \cong 15$
sources in $\approx 1.2$ days. Consequently, {\it GLAST} will scan the
entire sky to the level of $\phi_{-8} \cong 15$ in a time slight
longer than one day. For this, we can take the EGRET phase 1 catalog
as a guide to what {\it GLAST} will observe.

\begin{figure}
\begin{center}
\vskip-1.0in
\leavevmode
\hbox{%
\epsfxsize=3.0in
\epsffile{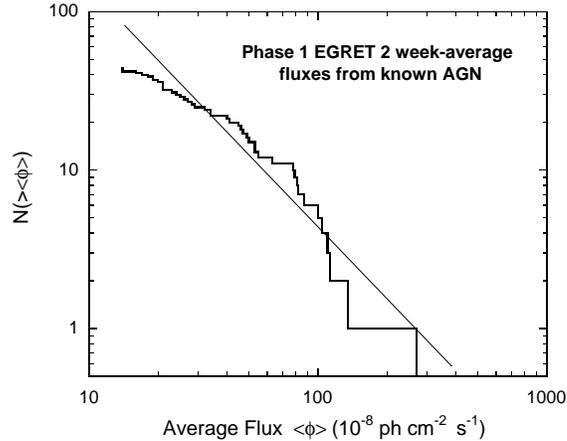}}
\caption{ The size distribution of blazar flares 
with 2-week--average fluxes $\langle \phi \rangle$ measured during the
phase 1 EGRET all-sky survey \cite{fic94}. The straight line
has slope $-3/2$.}
\label{fig1}
\end{center}
\end{figure}

\section{Comparison with EGRET Phase 1 Catalog}

Phase 1 of the {\it Compton Gamma Ray Observatory} observing program
took place during a 596 day interval from 1991 April 22 - 1992
November 17 \cite{fic94}.  Fig.\ 1 shows the size distribution of the
two-week average fluxes of radio-loud quasars and BL Lac objects
measured with EGRET during Phase 1, which consists of 44
high-confidence detections from 25 different blazars. Because the {\it
CGRO} Phase 1 was an all-sky survey, it will be less biased to reveal
the outcome of GLAST scanning observations than later EGRET catalogs
where the number of detections is biased by the many pointings in the
direction to previously known bright gamma-ray sources. Even so, the
Phase 1 exposure map of EGRET is far from uniform. One must keep in mind
that the two-week fluxes over shorter, brighter flaring phases, so that
the estimated blazar flaring rate could be increased if the duty cycle
of flares was better known cite{wal00}.

\begin{figure}
\begin{center}
\leavevmode
\hbox{%
\epsfxsize=3.5in
\epsffile{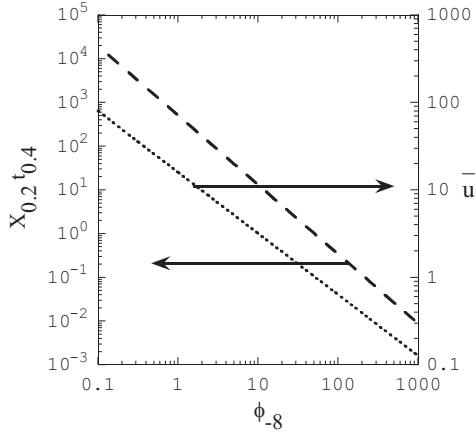}}
\caption{Minimum required observing time for {\it GLAST} in scanning mode
 to satisfy 5 count detection, 5 sigma sensitivity criteria for a
high-latitude source at the integral photon flux level
$10^{-8}\phi_{-8}$ ph($>100$ MeV) cm$^{-2}$ s$^{-1}$. Right axis is
$\bar u = \bar E/E_{100}$, the photon energy in units of 100 MeV that
satisfies these criteria in the shortest time. These results are for
sources with flat $\nu F_\nu$ spectra, i.e., $\alpha_\nu = 0$, and assumes
that only the extragalactic diffuse background contributes to 
background counts. The solution for $\bar u$ is valid for $1 \lesssim
\bar u \lesssim 100$.}
\label{fig2}
\end{center}
\end{figure}

The EGRET field-of-view was about 1/24th of the full sky, so that the
Phase 1 duration was about twice as long as the time required to
survey the entire sky with two-week pointings.  During this time,
EGRET detected some 3-10 sources at the level $\phi_{-8} \gtrsim
100$. Thus we expect {\it GLAST} to detect 2-5 such bright sources
during a one-day scan. Although uncertain in view of the small numbers of
blazars detected
at bright flux levels with EGRET, one
or two blazars brighter than $\phi_{-8} \gtrsim 200$ should be
detected with {\it GLAST} every 3-4 days. A comparable number of
unidentified $\gamma$-ray sources should also reach these levels
\cite{rb01}.


\section{Discussion}

Fig.\ 2 shows the amount of time in the scanning mode required for
{\it GLAST} to detect sources at given flux levels, using equations
(10) and (11). Once the source flux $\phi_{-8} \gtrsim 200$, dozens of
photon counts can be expected on a $10^4$ second timescale. Changing
to pointing mode for a non-Earth-occulted source will increase the
source counts by a factor of $\sim 3$-5, with the larger number
applying only to those few sources that are favorably located so that
Earth occultation is not severe.  At these flux levels, there are now
sufficiently many photons to perform time-resolved
spectroscopy. Sources exceeding the level of $\phi_{-8}
\approx 200$ must be carefully and quickly reviewed to determine
whether to issue an alert.  Coordinated sensitive radio, optical, UV,
and X-ray measurements would greatly strengthen tests of dominant
radiation processes and inferred locations of the $\gamma$-ray
emission sites.  At these bright flux levels, $\gamma\gamma$
absorption and spectral evolution can be studied to deduce Doppler
factors and monitor Doppler factor variations \cite{dg95,ds02}.

Fig.\ 2 also shows that {\it GLAST} in the scanning mode will be
sensitive to high-latitude $\gamma$-ray sources at the level of
$\phi_{-8} = 2$ and 0.2 over $\approx 2$ weeks and 2 years observing
time, respectively. Thus {\it GLAST} will reach flux levels of
$\approx 6\times 10^{-13}$ ergs cm$^{-2}$ s$^{-1}$, fifty times
fainter than EGRET's limiting sensitivity. As can be seen from Fig.\
2, good low-energy ($< 100$ MeV) response and calibration of the 
LAT on {\it GLAST} are important for identifying transients on the 
shortest time scales.

\vskip0.2in

We thank John Mattox for identifying an error in an early draft,
and acknowledge useful comments from Seth Digel.  The work of CDD is
supported by the Office of Naval Research and NASA {\it GLAST} science
investigation grant DPR \# S-15634-Y. The work of BLD is supported by
NASA {\it GLAST} Science Investigation grant NAG5-9712.


\begin{thebibliography}{99}

\bibitem[1]{lm83}
Li, P.-P., and Ma, Y.-Q. 1983, ApJ, 272, 317.
\bibitem[2]{glast} See LAT Instrument Performance at www-glast.stanford.edu.
\bibitem[3] {sre98} Sreekumar, P.~et al.\ 1998, ApJ, 494, 523 
\bibitem[4]{hks97}
Hunter, S. D., Kinzer, R. L., and Strong, A. W., 1997, in the 
Fourth Compton Symposium, ed. C. Dermer, M. Strickman, and J. Kurfess (AIP: New York), p. 192.
\bibitem[5]{fic94} Fichtel, C. E., et al. 1994, ApJS, 94, 55.
\bibitem[6]{wal00} Wallace, P.~M., 
Griffis, N.~J., Bertsch, D.~L., Hartman, R.~C., Thompson, D.~J., Kniffen, 
D.~A., \& Bloom, S.~D.\ 2000, ApJ, 540, 184 
\bibitem[7]{rb01} Reimer, O., and Bertsch, D.\ L.\ 2001, 
 Proc. 27th ICRC (Hamburg), 2001, pp.2566-2569 (astro-ph/0108348)
\bibitem[8]{dg95} Dermer, C.~D., and Gehrels, N.\ 1995, ApJ, 447, 103 
\bibitem[9]{ds02} Dermer, C.~D., and Schlickeiser, R.\ 2002, ApJ, 575, 667 
\end{thebibliography}
\end{document}